# Epidemic clones, oceanic gene pools and eco-LD in the free living marine pathogen *Vibrio parahaemolyticus*

Yujun Cui[1,2], Xianwei Yang[1,2], Xavier Didelot[3], Chenyi Guo[1], Dongfang Li[2], Yanfeng Yan[1], Yiquan Zhang[1], Yanting Yuan[2], Huanming Yang[2], Jian Wang[2], Jun Wang[2], Yajun Song[1], Dongsheng Zhou[1], Daniel Falush[4], Ruifu Yang[1,2]

1. State Key Laboratory of Pathogen and Biosecurity, Beijing Institute of Microbiology and Epidemiology, Beijing 100071, China

2. BGI-Shenzhen, Shenzhen 518083, China

3. Department of Infectious Disease Epidemiology, Imperial College, Norfolk Place, London W2 1PG, UK

4. Medical Microbiology and Infectious Diseases, College of Medicine, Swansea University, Institute of Life Science, 5th floor, Singleton Park, Swansea, SA2 8PP, Wales, United Kingdom

Y. C. and X. Y. contributed equally to this work; R. Y. and D. F. contributed equally to this work.

Corresponding authors: R. Y. (ruifuyang@gmail.com), D. F. (daniel_falush@eva.mpg.de), or D. Z. (dongshengzhou1977@gmail.com).

**Abstract**

We investigated global patterns of variation in 157 whole genome sequences of *Vibrio parahaemolyticus*, a free-living and seafood associated marine bacterium. Pandemic clones, responsible for recent outbreaks of gastroenteritis in humans have spread globally. However, there are oceanic gene pools, one located in the oceans surrounding Asia and another in the Mexican Gulf. Frequent recombination means that most isolates have acquired the genetic profile of their current location. We investigated the genetic structure in the Asian gene pool by calculating the effective population size in two different ways. Under standard neutral models, the two estimates should give similar answers but we found a 30-fold difference. We propose that this discrepancy is caused by the subdivision of the species into a hundred or more ecotypes which are maintained stably in the population. To investigate the genetic factors involved, we used 51 unrelated isolates to conduct a genome-wide scan for epistatically interacting loci. We found a single example of strong epistasis between distant genome regions. A majority of strains had a type VI secretion system associated with bacterial killing. The remaining strains had genes associated with biofilm formation and regulated by c-di-GMP signaling. All strains had one or other of the two systems and none of isolate had complete complements of both systems, although several strains had remnants. Further "top down" analysis of patterns of linkage disequilibrium within frequently recombining species will allow a detailed understanding of how selection acts to structure the pattern of variation within natural bacterial populations.

**Introduction**

Population genetics theory relates patterns of neutral genetic diversity to demography. Inference methods based on the theory have been applied to numerous organisms to make estimates of population sizes and how they change over time, e.g. (Li and Durbin 2011) and to detect barriers to dispersal and gene flow between regions, e.g. (Pritchard, et al. 2000). Such tools would seem to be particularly desirable for microbes since direct observation of patterns of dispersal in nature is extremely challenging. Population genetics also provides a null model for variation, allowing outlier loci that have been affected in particular ways by natural selection to be identified, e.g. (Voight, et al. 2006).

Most population genetic theory is based on the assumption that the variance in reproductive success of organisms is small relative to the number of individuals in the population (Eldon and Wakeley 2006). However this assumption is problematic for example for marine animals with planktonic larvae, where individuals can sire a significant proportion of the entire population (Eldon and Wakeley 2006). Bacteria typically divide by binary fission but fitness differences between lineages – whether due to genetics or local environment - can persist over many generations with the result that individual lineages can rise to a high frequency in the local or global population (Maynard Smith, et al. 1993). Bacteria belonging to these "epidemic clones" may be subject to recombination, meaning that successful lineages do diversify by recombining with other members of the population, but a sufficient proportion of each genome remains unrecombined that the clonal ancestral relationship – i.e., cellular parentage - is still evident (Croucher, et al. 2011; Didelot, et al. 2011). This means that estimates of effective population size are unlikely to relate in a straightforward way to the number of organisms in the species, either for planktonic organisms or prokaryotes.

Identifying barriers to the geographical dispersal of bacteria based on population genetic approaches is complicated by the fact that proximity may not be the most important factor in determining patterns of gene flow between strains. For example in *Campylobacter jejuni*, lineages have different host ranges and preferentially recombine with other lineages that share the same host (McCarthy, et al. 2007; Sheppard, et al. 2008). In *Escherichia coli* recombination is suppressed by DNA sequence mismatches between donor and recipient (Matic, et al. 1994). Recombination rates are higher between more closely related strains in *E. coli* (Didelot, et al. 2012; Shen and Huang 1986), which is broadly analogous to a systemic preference for inbreeding in eukaryotes and its population includes several 'phylogroups' which are each distributed globally (Tenaillon, et al. 2010).

Despite these issues, there have been some successful applications of population genetic approaches to investigate the demography of bacteria. For example, *Helicobacter pylori* colonizes the human stomach and does not survive long outside its host. Distinct populations of bacteria found in hosts in different geographic locations (Falush, et al. 2003). Populations that have likely been through a bottleneck, such as those carried by Maoris and Native

Americans harbor less diverse populations of bacteria. There is also evidence of genetic admixture between strains from distinct populations when they coexist within a human host population, e.g. due to human admixture (Yahara, et al. 2013).

The successful application of population genetic tools to *H. pylori* reflects a high rate of genetic exchange between strains, due to recombination during mixed infections of the same host (Kennemann, et al. 2011). This means that in particular geographic locations, there is reassortment of genetic variants in bacterial population such that variants on different parts of the chromosome are in approximate linkage equilibrium with each other. The bacteria spread relatively slowly, relative to their rate of recombination so that epidemic clones, if they exist at all, are uncommon locally and absent globally.

It is an open question as to whether similar approaches can be applied for prokaryotes that persist in the environment and therefore are not dependent on hosts to transport them from location to location. While there are many bacteria that show phylogeographic variation (Martiny, et al. 2006), this can be due to geographical adaptation and need not reflect barriers to dispersal. Even if there are geographical barriers, the organism may not recombine sufficiently frequently to break down genetic associations due to clonality in particular locations. A study in *Vibrio cholera* found evidence of geographic-based population structure in integrons, but detected none in the core genome, based on variation at six housekeeping genes (Boucher, et al. 2011). To our knowledge the twin criterion that hold for *H. pylori*, namely of an approximate absence of linkage disequilibrium within geographic populations and disequilibrium between populations due to divergence in gene frequencies at multiple loci, has not been shown to apply to any free-living bacteria.

*Vibrio parahaemolyticus* (VP) is an ocean bacterium that also causes infections of tens of thousands of people around the world each year (Su and Liu 2007; Yeung and Boor 2004). VP was first identified as the causative agent of a food poisoning outbreak in Japan in 1950 (Fujino, et al. 1965), and later found to be present in various environments, including ocean sediments, plankton and multiple types of sea-foods (Su and Liu 2007; Yeung and Boor 2004). Sporadic cases and small scale epidemics were observed worldwide and in 1996, a VP pandemic started in south east Asia, quickly spreading to Europe and America (Nair, et al. 2007). This pandemic was caused by VP serotype O3:K6 and its serovariants, all belonging to the same clonal group (Han, et al. 2008; Okuda, et al. 1997).

In this manuscript we first investigate the global population structure of the species. We used 157 genome sequences collected globally between 1951 and 2007 to demonstrate that despite the presence of epidemic clones, VP also exhibit segregation of variation between oceans consistent with conventional population genetic models of limited dispersal between gene pools and random assortment of variation within gene pools.

Having shown that the Asian ocean constitutes a single gene pool, analogous to a freely mating population in eukaryotes, we investigated the structure of genetic variation within it. We calculated the effective population size in two different ways. The gene pool effective population size reflects the number of lineages that contribute DNA to the gene pool of future generations, whether by recombination or direct descent. The genealogical effective population size reflects the number of clones that successfully contribute progeny to future generations. According to the standard assumptions of coalescent theory which include neutral evolution and random genetic exchange between individuals, the two approaches should give similar answers. However, our estimates of these two quantities differ by approximately a factor of 30.

We rule out a number of likely sources of the discrepancy, including statistical uncertainty and focus on the possible role of natural selection in maintaining multiple clonal lineages stably in the population. Such selection would require epistasis interacting at many different loci in order to structure the population into distinct genotypic niches. We find evidence for one such interaction. Further understanding of the source of the discrepancy should allow new insight into the forces that determine patterns of diversity in bacterial populations.

**Results and Discussion**

**Evolutionary history of VP**

We first used a Neighbor-Joining tree to explore the relationships amongst our strains using 327,904 high quality SNPs found in 157 genomes (Figure 1). This tree was star-like and many of the deep branches had small supporting bootstrap values (Figure S1), suggesting frequent recombination. Nonetheless, we observed 21 distinct groups each containing 2 to 35 closely related strains (Figure S1). In 13 of these groups, the strains were isolated from extended time periods and diverse locations (Table S1).

The presence of closely related strains in the sample allowed us to investigate evolutionary processes within a clonal context. We focused on the pandemic lineage (named CG1) and its relatives. This pandemic lineage included 35 strains with 1,106 SNPs differentiating them (Table S2). After excluding variation at sites likely to have been affected by recombination (dense SNP regions and homoplasies), the remaining 189 SNPs were used to construct a maximum likelihood tree (Figure 2A). Sublineages in this tree were present in multiple geographic locations, with the exception of one that was only found in Taiwan, implying very rapid spread of the bacteria. Superimposing the likely recombination sites onto branches of the phylogeny revealed that they were concentrated in three branches of the phylogeny and within a 158.5 kb region (182820-341273 in chromosome I) surrounding the O- and K-antigen encoding gene cluster which led to serovar shift events (Figure 2A and B). We designated this stretch of sequence as being a putative recombination "hot region".

One strain (S093) and one clonal group (CG2) were closely related to CG1 and this was supported by high bootstrap values (Figure 1 and Figure S1). The strain S093 had 2,323 SNPs separating it from CG1, most of which were concentrated in 13 genomic regions (Figure 2C). These dense SNPs regions are almost certainly caused by homologous recombination events. To investigate the SNP distribution amongst less closely related bacteria, we plotted the same distribution for all strains from CG1 and CG2 plus S093 (Figure 2D). Most regions of the genome contained a high frequency of SNPs but there were also some long regions of low SNPs density. These putatively represent regions where the strains share a common clonal frame, i.e., they have all inherited DNA by vertical descent from the same ancestor in these regions. When an additional randomly selection strain was added, the distribution of SNPs around the genome became more even and regions of low SNPs density were very short (Figure 2E). This result is consistent with an absence of shared clonal frame and with the additional strain being an unrelated member of the same frequently recombining bacterial species.

**Global population structure of VP**

Frequent recombination at the species level was confirmed by measures of linkage disequilibrium, which we compared with that in other bacterial species (Figure S2). Linkage disequilibrium in VP decays to less than half of its initial value within 250bp. In this comparison, both *H. pylori* and *E. coli* had faster initial decays than VP. The former is presumably a consequence of extensive recombination during mixed infections of the same human stomach (Didelot, et al. 2013) combined with the breaking up of imported fragments into short fragments during incorporation into the recipient genome (Kulick, et al. 2008). However, the LD between SNPs on distant parts of the VP chromosomes was lower than for any of the other species.

In order to characterize patterns of gene flow, it is necessary to take appropriate account of shared ancestry due to clonal relatedness, which may have a significant impact even if the shared clonal frame represents only a small fraction of the genome. Methods for quantifying gene flow also need to take into account that DNA is imported in segments, so that markers that are close to each other on the chromosome have a shared history. Phylogenetic methods ignore physical linkage and treat SNPs as providing independent information and therefore can over-estimate the statistical confidence in particular subdivisions.

We have therefore supplemented phylogenetic analysis with an analysis based on chromosome painting (Lawson, et al. 2012). In a first step, each isolate is treated as a "recipient". Its chromosomes are "painted" as a series of segments from the other "donor" strains in the sample. Each donor is the most closely related strain to the recipient for that stretch of chromosome. Fragment boundaries indicate changes in ancestry so that each fragment can be thought as being inherited as a distinct unit from the gene pool. The number of fragments that a particular donor contributes to the recipient in this painting process is called their pairwise coancestry value. In a second step, fineSTRUCTURE is used to group together strains with statistically indistinguishable coancestry values into populations. The approach provides a statistically rigorous method of establishing whether different strains have distinct patterns of shared coancestry (Lawson and Falush 2012).

Before applying fineSTRUCTURE, the dataset was thinned down to 71 strains (Table S1) by selecting one representative strain from each of the clonal groups with average nucleotide identity above 99.97% (compared to an average of about 98.44% across the dataset).

fineSTRUCTURE identified 11 populations (Figure 3A). The subdivisions found were completely concordant with those found by neighbor joining tree, with the exception that a single isolate, S058, placed in the Asia-pop 1 population by fineSTRUCTURE was clustered into Hyb-pop 2 in Figure 1. The largest Asian sub-population, Asia-pop 1 had 44 members, who were painted with an average of 2,224 chunks of mean size 1.8 Kb. By contrast, the three isolates in this analysis from Asia-pop 6 were painted using far fewer chunks (451 on average). These three isolates had long stretches of high similarity reflecting clonal descent and each one was painted using large chunks from the other Asia-pop 6 isolates. Asia-pop1, as well as containing the largest number of strains, had the lowest coancestry values between members of the same population.

We hypothesized that Asia-pop 6 and many of the other smaller populations identified by fineSTRUCTURE in this analysis do not represent distinct, freely recombining gene pools but rather sets of strains that share DNA by direct clonal descent. To test this hypothesis, we performed analyses with a dataset consisting of all isolates from Asia-pop 1 and one isolate each from the other populations. If the other populations consist of clonally related strains sampled from a single Asian gene pool and no further cryptic signals of clonality remained, then the representatives of these populations should be absorbed into the Asia-pop 1 population in this new analysis.

In this new analysis, we found four populations, one of which contained the isolates from Asia-pops 1-8, while the other three populations consisted of the single representatives from US-pop 1, Hyb-pop 1 and Hyb-pop 2 that were included in the analysis (Figure 3B). Our interpretation is that the strains in Asia-pops 1-8 come from a single gene pool and the different strains in the population are clonally related, while the other populations consist of strains with distinct ancestry and patterns of gene flow. We also estimated a NJ tree for the same set of strains and found that all strains of Asia-pops are approximately equally related with each other, except one strain, S058, in Asia-pop 1 was clustered with the strain of Hyb-pop 2 to form a distinct lineage, as well as the US-pop 1 and Hyb-pop 1 formed the other distinct lineages (Figure S3).

Support for the hypothesis of our sample containing 51 isolates from a single freely recombining Asian gene pool is provided by estimates of the value of linkage disequilbrium $r^2$ between these strains. The expectation of $r^2$ is dependent on the sample size as well as the linkage disequilibrium in the population they are sampled, and in the absence of any LD, its expectation is 0.0196 is predicted for 51 unrelated isolates (Park 2012). The estimated value of 0.02 for SNPs greater than 1 kb apart is very close to this expectation and the variance between subsets of 5 isolates from this 51 is very low (Figure S4A), consistent with them all being sampled independently from the same population. A much less good fit to the theoretical expectation is found when all 71 isolates from the thinned dataset are used in an equivalent analysis (Figure S4B), confirming that the reduction to 51 isolates has eliminated most or all of the clonal structure present in the dataset.

US-pop 1 was the most distinct, while the hybrid populations showed elevated coancestry with each other and with US-pop 1 compared to members of the single inferred Asia population. Strains from Hyb-pop 1 and Hyb-pop 2 had intermediate ancestry between US-pop 1 and Asia-pop 1 but were also related to each other. Hyb-pop 1 and Hyb-pop 2 were provisionally designated as hybrids based on their intermediate ancestry patterns, but the pattern is also consistent with them being representatives of other distinct gene pools. In contrast to some previous observations in other bacteria (Shapiro, et al. 2012), We did not find evidence for systematic differences in the flexible genome between populations (Figure S5). As has previously been found in *H. pylori*, similarity in flexible gene content is disproportionately determined by large gain and loss events and as a result provides less phylogeographic signal than core genome polymorphisms (Gressmann, et al. 2005).

The conclusions based on the coancestry matrix and Neighbor-Joining method were confirmed by a table of average nucleotide diversity within and between populations (Table S3). Members of the Asia populations are more closely related to each other on average (37,669 SNPs in the core genome distinguishing pairs of strains or ~1% sequence divergence) than the members of US-pop 1 (41,036 SNPs), with Hyb-pop 1 and Hyb-pop 2 displaying intermediate values.

**Oceanic distribution of the VP populations**

In order to investigate the geographical distribution of the populations, we studied the relationship of these strains with the larger number that have been genotyped by MLST. We first used the linkage model of STRUCTURE to identify populations within the MLST data (Figure 4A). Since data from 7 loci may not be sufficient to describe genomic patterns of relatedness, we considered the populations identified by STRUCTURE to be valid only if they were concordant with those found using the genomic data. Based on this criterion, two populations were validated, with suggestive evidence for a third. High ancestry from the R1 population (orange) in the STRUCTURE analysis correlated with the strains having high coancestry with the three members of the US-pop 1 at the MLST loci (Figure 4B). Strains with high ancestry from R2 (red) belonged to the population containing the epidemic clone (Asia-pop 6). 66 of the 80 isolates from the Mexican Gulf were from R1, which were also found at lower frequencies in other US locations but were almost entirely absent elsewhere in the world (Figure 4C). R3 correlated with Hyb-pop 1 which was found in the pacific. This geographic pattern provides suggestive evidence that Hyb-pop 1 constitutes a distinct geographical gene pool.

Since no group corresponding to Hyb-pop 2 was identified in the MLST data, our knowledge of their distribution remains limited to the 5 genomes in our sample, which were found in Europe, Asia and America. This is consistent with the population representing a clonal lineage that had spread between geographic areas and had differentiated via import of local DNA but additional data would be necessary to confirm this hypothesis.

**Genetic structure of the Asian population**

Since Asian VP represents a single freely mixing population, population genetic methods can be used to estimate $N_e r$, the product of the effective population size $N_e$ with the per-generation rate $r$ at which sites recombine. This is a natural parameter for a coalescent model where recombination drives diversification, and its estimation can be approached in two ways.

A first approach is to measure the effective population size experienced by individual sites in the genome. This is the approach conventionally used to measure effective population size in eukaryotes. The larger the effective population size, the longer the time to the expected most recent common ancestor at each site and the greater the opportunity for diversity to be accumulated by mutation. We base our estimate of the population size on the average diversity over a large number of sites. Different sites have different genealogies due to frequent recombination (Figure 5A), so this measure reflects a large number of independent coalescent events. Under a coalescent model with any amount of recombination, the expected evolutionary distance between two genomes at a given site is $\pi = 2N_e\mu$ where $\mu$ is the per-site per-generation mutation rate (Wang 2005). In order to try and measure the effective population size for unselected sites, the most appropriate value of $\pi$ for this calculation is the pairwise synonymous distance between 51 unrelated strains in the Asian population (44 strains from Asia-pop 1 and one strain each from the other Asia-pops), which is $\pi = 0.0258$. We inferred the $r/\mu$ (ratio of size of recombination regions against the number of mutation sites) from strains of clonal groups, which given the $r/\mu = 757$ (Figure 6A). Since $N_e r = (N_e\mu)(r/\mu) = (\pi/2)(r/\mu)$, this approach leads to $N_e r = 9.8$. Multiplying $r/\mu$ (Figure 6A) with $\pi$ produces an estimate of the relative effect of recombination and mutation $r/m = 19.6$ (Jolley, et al. 2005) which is slightly higher than found in a previous MLST study ($r/m = 16$)

(Yan, et al. 2011).

An alternative approach is to measure the effective population size for the organism as a whole, based on the rate of genealogical coalescence of strains in the sample. Since recombination occurs progressively over time, the proportion of shared clonal frame between strains provides information on the age of their common ancestor. We can therefore use pairwise genetic divergence between strains to estimate coalescence times. These coalescence times can then be compared with the distribution predicted by coalescent theory in order to estimate the effective population size.

We assume that genetic divergence within clonal complexes mostly occurs by recombination (as suggested by the above estimate $r/m$=19.6). Bacteria in the Asian populations with no trace of clonal relatedness have a median of $d_{unrelated} = 37,713$ SNPs distinguishing them in the rest of the core genome (Figure S6). As strains diverge by recombination, the number of SNPs distinguishing them will increase at a steadily decreasing rate, until it approaches this value. The probability that a site is unaffected by recombination for two strains that share a common ancestor $T$ generations ago is $\exp(-2rT)$ so that their expected number of differences is $d_{unrelated}(1 - \exp(-2rT))$.

Theory predicts that the rate of coalescence is proportional to effective population size.

Specifically, in a coalescent tree, the expected time $T$ until $n$ individuals have $m$ ancestors is equal to $2N_e \left(\frac{1}{m} - \frac{1}{n}\right)$. Combining this formula with the estimate for divergence between strains as a function of times above give the expected distance corresponding to the $m^{th}$ common ancestor as being $d_{unrelated} \left(1 - \exp\left(-4N_e r \left(\frac{1}{m} - \frac{1}{n}\right)\right)\right)$ which is a function of $N_e r$ and $n$, the number of samples from the population. In order to estimate $N_e r$ based on this formula, the expected distances between coalescences were compared to the ones estimated using the UPGMA algorithm applied to the matrix of pairwise distances in the Asian data (Figure 6B).

In practice, our strains are not sampled independently from a homogeneous population. Our sampling is very uneven, especially because of the bias towards disease causing isolates. Epidemic clones can also skew the distribution away from coalescent expectations. These factors mean that the effective number of independent samples from the population, $n$ may be smaller than the number of isolates (Fraser, et al. 2005). In fact, by minimizing the sum of least square errors though exhaustive searching for the optimal parameters in the bi-dimensional space with $n$ ranging from 10 to 145, and $N_e r$ from 10 to 500, we found that the best fit to coalescent expectations was for $n = 60$ and $N_e r = 268$ (Figure 6B). This result implies that only 60 of our 145 strains from the Asian population can be considered representative samples of the non-epidemic population structure. The other strains represent repeat samplings from the population. We also evaluated other values of $N_e r$ that from 10 to 450 in the whole-genome approach, to observe the best fit with the observed values (Figure 6B). The results show that lower $N_e r$ values predict very different coalescence rates to those observed, implying $N_e r$ is very unlikely to be below 150.

The two approaches above to estimate $N_e r$ give the strikingly different values of 9.8 and 268 for the site-by-site and the whole-genome methods respectively. Assuming a high recombination rate *r* of the order of $10^{-5}$ per generation as in *H. pylori* (Didelot, et al. 2013; Kennemann, et al. 2011), the effective population size $N_e$ would be $10^6$-$10^8$ which is compatible with previous estimates (Fraser, et al. 2009). The fact that the two estimates differ by a factor approximate 30 implies that genetic drift is stronger when measured at individual sites than for whole organisms. In eukaryotes, recombination normally increases the effective population size at particular sites by reducing the effect of selective sweeps and background selection at all but the nearest sites (Charlesworth 2009), and this relationship was also assumed to hold in bacteria (Castillo-Ramírez, et al. 2011). Our observation, however, implies that recombination can have the opposite effect in VP. Specifically, if (hypothetically) there were sites in the genome that never recombined, the expected diversity of these sites would be higher than the rest of the genome.

**Ecotypes as an explanation for the different $N_e r$ estimates**

Under the population size estimates implied by the nucleotide model we expect only a small number of approximately unrelated lineages (>32,000 nucleotide differences between strains) to be observed in the population, i.e., we would expect on average 13 based on a sample of 60 or 14 based on a sample of 145 or 15 based on sampling the whole population. In fact, we observe more than 50 in our sample and the coalescent estimate we obtain suggests that there are hundreds in the population as a whole.

Our observation could be explained by selection at the organismal level that prevents individual clones from becoming numerically dominant. This might be caused by phages targeting common clones, or the partitioning of the VP population into ecologically distinct niches, for example caused by adaptation to living on organisms or particles of particular sizes (Shapiro, et al. 2012). This selection would maintain multiple ecotypes stably in the population and therefore reduce the rate of coalescence observed between strains with different ecotypes (Figure 5B), but would not necessarily prevent variants outside the ecotypically selected regions sweeping frequently to fixation in the gene pool (Charlesworth 2006) and thus could be consistent with the higher effective population size for clonal coalescence than for nucleotides.

Since our estimates suggest the maintenance of several hundred lineages independently in the population, it suggests that niche structure is maintained by multiple phenotypic traits. Furthermore, the absence of easily detectable barriers to gene flow between lineages created by any of these traits suggests that the ecotype structure may be stable over time and not a precursor to genomic divergence and speciation as suggested by most previous discussion of ecotype models (Achtman and Wagner 2008; Cohan and Koeppel 2008; Shapiro, et al. 2012).

If the species is partitioned into multiple distinct niches, then this should be detectable by non-random associations of alleles that are relevant to particular niches. In order to minimize the possible effect of population structure, we tested for associations between 271,945 biallelic SNPs in a sample of 51 unrelated isolates in the Asian population, using a Fisher exact test. Since our approach does not make any assumptions about what the adaptive phenotypes might be or how it is partitioned amongst isolates, it is distinct from and complementary to that of Shapiro et al. who attempt to identify ecologically differentiated loci based on a particular niche partitioning (Shapiro, et al. 2012) and can be considered a completely top-down approach to identifying ecological differentiation.

The great majority (96%) of associations with low $P$ value (less than $10^{-6}$) occur between sites that are less than 1 kb apart, reflecting LD at the level of individual genes. We used a Q-Q plot to investigate how many of the associations are genuine (Figure 7A). It can be seen that the values including linked sites starts to diverge substantially from the curve for only "unlinked" sites (>3 kb apart) at a $P$ value of approximately $10^{-6}$. The Q-Q plot for all sites becomes approximately linear at around $P = 10^{-8}$. The great majority of these are true associations due to linkage disequilibrium generated by co-inheritance of closely linked sites in recombination events.

Amongst the unlinked sites, the curve does not reach the same slope, and there is only one set of SNPs (LS001) that has $P$ values below $10^{-9}$ (Table S4). The SNPs are located in two regions that are 400 kb apart in the reference genome (Figure 8). In the other genomes sequenced here, the two regions are also located inside different scaffolds, implying that this association is not caused by physical linkage of the chromosome. Since the LD between these SNPs cannot be attributed to either population structure or to physical linkage, we hypothesize that it is caused by ecological factors and term it eco-LD, for ecological linkage disequilibrium.

One of the regions contained three SNPs in a hypothetical gene (VPA1081) that carried a transmembrane domain, and most of the SNPs (38 in 46) in the other region are located in two genes that code for LuxR family transcriptional regulator, CpsQ and CpsS, and the intergenic region between them. Both CpsQ (VPA1446) and CpsS (VPA1447) are involved in modulating cyclic dimeric GMP (c-di-GMP) signaling and biofilm formation (Ferreira, et al. 2012).

Strong LD was also observed between alleles of SNPs and presence/absence of flexible genome blocks. Amongst a total of 8.04 Mb length of flexible genome blocks in 51 unrelated strains, 35 unlinked blocks with length of 114 kb that revealed strong association ($P < 10^{-6}$, indicated by inflection point of the QQ plot in Figure 7B). Interestingly, majority of these blocks (85/114 kb, 75%) were associated with LS001 (Table S5 and Figure 8), the only set of SNPs that were identified with real epistasis association. Two major functional units were annotated on these blocks, one encoded type VI secretion system 1 (T6SS1, Table S6), which is up-regulated and has anti-bacterial activity under warm marine-like conditions (Salomon, et al. 2013). Another encoded a group of cellulose biosynthesis-related proteins, which were important for biofilm formation and regulated by c-di-GMP signaling, hence functionally close related with CpsQ and CpsS in LS001 (Tischler and Camilli 2004). The c-di-GMP-mediated decision-making network influences the switch between two different social behaviors of VP: staying put and forming a structured biofilm on the surfaces, and spreading over the surfaces by swarming motility (Gomelsky 2012). Therefore, the possible epistasis signal revealed here might have effects on the fitness of VP in different environments that make up parts of its overall niche, e.g., on particular hosts, surfaces or in the open sea.

To further investigate the association across these epistasis loci, we examined the SNPs combinations in LS001 and the distribution of the associated flexible blocks for all 157 strains in our dataset. Two major groups could be defined by these SNPs, with most of cellulose synthesis-related genes were present in EG1 isolates, whereas T6SS1 encoding genes were exclusively distributed in EG2 isolates (Figure S7). Interestingly, EG2 contained overwhelming number of clinical isolates (92.5%, Figure S7), and the chi-square test for the unbalanced distribution of clinical isolates in two groups is significant ($x^2$= 34.1438, $P <$ 0.0001, but $P$ = 0.052 if only one strain from each clonal group is used in the analysis), suggesting that the epistasis sites may influence infection ability or virulence, and was consistent with previously observation that T6SS1 is predominantly present in clinical isolates (Yu, et al. 2012).

There are no specific geographic or temporal structure that associated with dividing of the two groups (Figure S7), and although the nucleotide polymorphisms within each group are mostly in linkage disequilibrium with the epistatic sites, the presence-absence regions associated with EG1 and EG2 both show similar genetic diversity compared with the whole core genome ($\pi$ = 0.0091, 0.0089 and 0.0092 for EG1, EG2 and core genome, separately) and a star-like pattern of variation within them (Figure S8), consistent with frequent recombination reasserting variation within these regions. This pattern of variation within the genes is consistent with the reassembly of these co-adapted gene complexes by natural selection on multiple occasions.

The fact that we have found only a single convincing example of epistasis in this screen does not necessarily imply that epistasis is rare, since our statistical power is limited based on 51 unrelated strains, mainly due to the very large number of possible pairwise associations. It does suggest that strong epistasis between high frequency alleles is rare and that if there are distinct, stably maintained ecotypes within the Asian gene pool, they are maintained either by single genomic loci with multiple interacting epistatic partners or by quantitative traits due to multiple loci or by epistatic interactions that are in other ways too complex to be assayed effectively by our scan.

**Alternative explanations for differences between the two $N_e r$ estimates.**

While niche differentiation provides a sufficient and biologically compelling explanation for the very different estimates of $N_e r$ generated by organismal and nucleotide based approaches, there are a number of other factors that could cause the two estimates to differ. We have investigated several and found that none of them is sufficient to explain the discrepancy on their own.

Firstly, the disparity between the two estimates of $N_e r$ might in principle be explained by an under estimation of $r/\mu$ or equivalently $r/m$, which would happen if polymorphisms treated as mutations in Figure 6A were in fact introduced by recombination. This, however, seems unlikely since value of $dN/dS$ for these polymorphisms is 0.98, which is very close to the value expected for new mutations of 1.0 (Rocha, et al. 2006) and very different from the value found in our data of 0.08 for $dN/dS$ for polymorphisms introduced by recombination events. We also employed another independent recombination detection method, RecHMM (Zhou, et al. 2014), to infer the recombination events and although the result showed a wider recombination region ($r/\mu = 1,117$, Figure S9), the $N_e r$ of site-by-site is 14.4, which is still far smaller than the value based on clonal coalescence.

Secondly, to evaluate statistical uncertainty, we calculated the confidential interval for each estimate. For the site-by-site approach, the $r/\mu$ was inferred by slope of linear regression in Figure 6A. The 99% CI of the slope was [402, 1111], which given the upper limit of the $N_e r$ by individual site as 14. Then we inferred the CI of $N_e r$ for the whole organism by two different ways. Firstly, we random selected 50% of SNPs from the whole dataset to calculate the $N_e r$ and repeat the process for 1,000 times. The results generated an approximate normal distribution of the $N_e r$ and the 99% CI was [264, 272] (Figure S10A). The second way is to calculate the variance of the median of genetic distance across the isolates of Asia-pops, and then replaced them separately into the formula to acquire the CI. To cover the 99% of the genetic distance between pairs of isolates, the upper and lower limits of genetic distance were from 38,852 to 36,797, which generated the 99% CI of $N_e r$ as [245, 287] (Figure S10B). The lower limit of the $N_e r$ for the whole-genome (245) is still much larger than the upper limit of $N_e r$ for the individual site (14), indicating the conclusion of disparity between two estimates is robust.

Thirdly, as the strains were collected over a long time period, we considered that whether temporal sampling might have affected coalescence rate estimates. We split the dataset into old (1984-1995) and young (1996-2007) groups, and estimate the $N_e r$ in two different approaches for each. The $N_e r$ based on individual site were 10.22 and 9.88, and the values based on whole organism were 267 and 211, for old and young group, separately. Therefore temporal sampling effects cannot explain the conclusions.

Finally, we considered the possible influence to the $N_e r$ estimates by migration events between Asia-pops and other populations. Under a neutral population genetic model, the effect of population subdivision should be the same for nucleotides as the clonal tree. Therefore, this factor will not cause a bias based on random sampling. Since $F_{st}$ is low (see below), the nucleotide diversity estimates for the global population are similar to those of just the Asia-pops. Furthermore, the expected number of migration events during the observable period of clonal evolution is very small (see below). Therefore we do not think that this factor should cause a substantial bias in estimates.

**Migration and clonal divergence**

We can use standard population genetic theory estimate migration rates between oceans based on estimates the effective population size and of divergence between populations. This analysis should be considered with caution because we have shown above that the assumptions of equilibrium under a standard demic diffusion model are likely to be very unrealistic for the species.

The fixation index $F_{st}$ between 61 Asian isolates and the 3 strains from US-pop 1 used in the fineSTRUCTURE analysis was equal to 0.071. While Hyb-pop 1 revealed similar $F_{st}$ values with Asia-pop (0.021) and US-pop (0.027), Hyb-pop 2 seemed much closer with Asia-pop (0.027) but distinct with the US-pop (0.073), which was consistent with co-ancestral relationship inferred by fineSTRUCTURE analysis (Figure 3A).

Under standard population genetic demic diffusion model, the product of the effective population $N_e$ and the per generation migration rate $m$ is $N_e m = 6.6$ (Whitlock and McCauley 1999). Given the high rate of recombination, strains quickly acquire genes from their new local gene pool. Specifically, the clonal frame is mostly erased in the time period in which 60 strains coalesce to 50 ancestors (Figure 6B). The number of migration events per strain expected during that time is $2N_e m \left(\frac{1}{50} - \frac{1}{60}\right) = 0.05$. Most isolates from Asia were assigned to the Asian population which is consistent with this expectation. By contrast, the recent global spread of some epidemic clones is unlikely under the above neutral model. As noted above, this suggests that strong selection has been responsible for their rapid dispersal.

**Conclusions**

We found an order of magnitude difference between the effective population sizes for nucleotides in the genome and for organisms. This implies that standard population genetic theory, which predicts the two values should be the same, is not a good approximation for large bacterial populations. One possible explanation for the discrepancy is that the species is divided into multiple distinct ecotypes that coexist stably in the same population. We probed this explanation – which suggests that there are more than 100 distinct ecotypes - by looking for pairwise associations amongst loci. We found only a single strong example of epistasis, which is associated with how the cell attempts to modify its immediate environment, either via biofilm formation or killing other bacteria. Understanding the factors responsible for the large deviation from neutral populations should help us to gain considerable insight into fitness landscapes and patterns of recombination in natural bacterial populations.

Clones of VP can spread globally as illustrated for example by the distribution of the recently arisen pandemic lineage. Nevertheless, there are differentiated oceanic gene pools and the majority of strains have mosaic ancestry from their local gene pool. We have shown that the Mexican Gulf represents a gene pool distinct from that found in Asia. There are also hints of a distinct gene pool in the US pacific. Currently, we do not have sufficient data from genomes outside Asia to establish the number of distinct gene pools globally and the pattern of gene flow between them. Additional genomes will establish this and whether there are also adaptive differences in genome content between oceans.

## Materials and methods

**Bacteria selection**

A total of 156 strains of *V. parahaemolyticus* were selected for sequencing in this study (Table S1). These strains were isolated from human clinical cases (112), sea food (31) and environment (13), which came from 13 countries in Asia, Europe and North America, and the isolation time were between 1951 and 2007. All 156 strains were involved in our previous microarray and MLST assays (Han, et al. 2008; Yan, et al. 2011). Genomic comparisons were performed together with the genome of RIMD 2210633 (accession number: NC_004603 for chromosome 1 and NC_004605 for chromosome 2), which were downloaded from the NCBI database (ftp://www.ncbi.nlm.nih.gov/). 35 of the 157 strains have previously been defined as pandemic strains (Han, et al. 2008; Yan, et al. 2011).

Bacteria were grown in the LB-2% NaCl agar at 37 °C, and the extraction of genomic DNA was performed by the classical phenol/chloroform method.

**Sequencing, assembly and annotation of coding sequences (CDS)**

Whole genome sequencing was performed using the Illumina Genome Analyzer II (Illumina Inc. U.S.A). The multiplexed paired-end libraries with an average insert size of 500 bp were constructed following the manufacturer's instruction, and the pair-end read lengths of 10 strains were 44 bp, two strains were 75 bp, and the remaining 144 strains were 100 bp. After trimming the adaptor sequences and removing the low quality reads, we obtained 81 Gb of high quality sequence data in total, corresponding to an average 92 fold coverage (effective depth) for each strain. The short reads were assembled using SOAP *denovo* (version 1.1.2) (Li, et al. 2010), which resulted in the contigs that average total length of 5.1 Mb (N50 = 21.3 kb) for each strain (Table S7). The CDS were predicted for each sequenced genome by using Prodigal (Hyatt, et al. 2010). Then the functional annotation of the amino acid sequences of predicted CDSs was performed by alignment against the non-redundant database of NCBI using BLASTp with the criterion of e-value < 1e-5, identity% > 40% and length coverage of gene > 80%. The functional domain of the hypothetical gene in LS001 was annotated by online alignment using SMART in EMBL (Letunic, et al. 2014) and CDD in NCBI (Marchler-Bauer, et al. 2013).

**Construction of the core and flexible genome**

We constructed the core-genome by comparing genome sequences of 157 VP strains to identify genomic contents shared by all of them. The 156 assembled genomes were aligned against the reference genome sequence (RIMD2210633) using BLASTn, to delineate shared genome regions with identity $\geqslant$ 90% and e-value < 1e-5. The defined core-genome of VP was consisted of 1,333 blocks 4.07 Mb in total length. After removing the regions that mapped to the core-genome, the remaining sequences for each strain were combined to obtain a set of strain specific sequences with redundancy. These redundant sequences were compared to each other using BLAT and grouped into 13,194 sets of sequences with identity $\geqslant$ 90% and match length $\geqslant$ 85% in pairwise comparisons. Then the longest sequence in each group of similar sequences was considered as representative for each group. This set of non-redundant sequences were merged and considered as the flexible genome for all 157 VP strains, with total length of 13.26 Mb. Furthermore, to perform the epistasis analysis, we constructed the flexible genome of 51 unrelated strains in Asia-population with length of 8.04 Mb, by using a same pipeline.

**SNPs identification**

We firstly aligned the contigs of each of 156 *V. parahaemolyticus* against the reference genome (RIMD 2210633) using MUMmer v3.20 (Delcher, et al. 2003) to obtain all potential SNP loci, and then filtered unreliable loci that were covered by less than ten effective supported reads or located in repeat regions. The effective supported read for one nucleotide was required to have a quality score greater than 20 and to not be located within 5 bp of one of the edges of the read. The repeat regions were defined in RIMD 2210633 genome according to the method described in previous research (Table S8) (Cui, et al. 2013). This included (1) Variable number tandem repeats identified by TRF 4.04; (2) Dispersed repetitive sequences found by BLASTn search of the RIMD 2210633 genome against itself with >95% identity and >50bp length; and (3) CRISPRs (clustered regularly interspaced short palindromic repeats) recorded in the CRISPRdb database (Grissa, et al. 2007).

Finally, we obtained 327,904 high quality SNPs, consisting of 84,101 nonsynonymous SNPs (nsSNPs), 540 of which resulted in premature stop codons; 208,086 synonymous SNPs (sSNPs) and 35,717 SNPs in intergenic regions.

**Inference of recombination sites within clonal groups**

Homologous recombination events were detected based on two types of signals: (1) regions of dense SNPs in a genome when compared with its close phylogenetic relatives and (2) homoplasies where the same allele was found in strains from different lineages which can not be attributed to inheritance from the same common ancestor.

Accordingly, we first detected regions contained dense SNPs by the method similar with that describe previously (Croucher, et al. 2011). Briefly, we constructed a neighbor-joining tree for each clonal group and determined the SNPs occurring on each branch using baseml in PAML software package (Yang 2007). Assuming neutrality and no recombination, the observed SNPs could be modeled as a binomial distribution with the given chromosome size and mean number of SNP per base (null hypothesis). Using a sliding window method, we checked whether the SNPs density in each genomic region followed this distribution ($P < 0.05$), and if the null hypothesis was rejected for a particular window, all SNPs in this window were defined as introducing by recombination. To reduce the false negative rate, we set different window sizes with 1,000 and 2,000 bp in independent runs, separately, and using each SNP as the beginning bound of the window, the moving window would traverse every SNP according to their order in the reference genome (equivalent to step size = 1 bp). All possible recombination sites detected in different runs were combined together to estimate the range of recombination regions, which was defined as containing recombed sites that all pair-wise distances between adjoined SNPs were less than 2,000 bp. The boarders of recombination region were defined according to the most distant SNPs in the identified dense SNP region. As the above definition of the borders would potentially reduce the real size of recombination segments, we also detected the clusters of SNPs in strains of clonal groups by using the software RecHMM (Zhou, et al. 2014), which generated comparable results to the wildly applied software ClonalFrame (Didelot and Falush 2007) but with much higher computational efficiency. The information of SNPs occurred on each branch of the phylogenetic tree was used as input file, and then we inferred the recombination regions by running RecHMM under the default parameters. After excluding regions with dense SNPs, we rebuilt the phylogeny and again using baseml to infer the SNPs on each branch. If the same SNP occurred in different branches, it was considered to be homoplasic.

**Construction of the phylogenetic tree**

The Neighbor-Joining trees were built using TreeBest software based on concatenated SNPs (http://treesoft.sourceforge.net/treebest.shtml). The maximum likelihood tree of 35 pandemic strains on the basis of 189 SNPs was constructed using PHYML with HKY model (Guindon and Gascuel 2003).

**Calculation of linkage disequilibrium (LD) decay**

Firstly, we identified SNPs from the published genome sequences of six other different bacterial species (Table S9) according to the SNP calling process described above. Then we calculated the value of $r^2$ by using Haploview software based on SNP sets (Barrett, et al. 2005). The main option was "-maxdistance 10 -minMAF 0 -hwcutoff 0" which meant the maximum inter-marker distance for LD comparisons was 10 kbp, the threshold of the minor allele frequency and the Hardy-Weinberg *P* value were set to 0 to include all the identified SNPs in the calculation.

To infer the initial LD decay rate among different species, we fitted the LD decay curves with the exponential function $y(x) = A + B \cdot e^{-x/x_0}$, where $x_0$ determined the decay rate of the curve, with a larger number indicating slower decay (Donati, et al. 2010).

**STRUCTURE analysis based on the published MLST data.**

The sequences of 7 gene fragments from 281 STs of VP were downloaded from the pubMLST database (http://pubmlst.org/vparahaemolyticus/). The software STRUCTURE (v2.3.2) with admixture model (Falush, et al. 2007) was used to conduct the population assignment based on this data as previously described (Falush, et al. 2003; Moodley, et al. 2009). The length of MCMC chain was set to 50,000 and the first 20,000 iterations were discarded as burn-in. The parameter K (number of population) was set from 2 to 20 and independent runs were conducted. The result for K=10, which had the highest marginal likelihood, was displayed by the software DISTRUCT and used in further analysis (Rosenberg 2004).

**Chromosome painting and fineSTRUCTURE analysis**

We used fineSTRUCTURE to infer the population structure of *V. parahaemolyticus* based on genome wide SNPs (Lawson, et al. 2012). We prepared the recombination map file assuming a uniform recombination rate for each SNP site as $1/(100 \times \text{alignment size})$. Independent runs of ChromoPainter were performed with different values of the parameter $k$ (100, 200, 500, 800, 1000, 1500, 2000), and as the value of the parameter $c$ tended to be constant when $k \geq 500$, we selected $k = 500$ for further calculation. The chromosome painting was conducted for each chromosome separately and the output files were integrated using ChromoCombine. For the fineSTRUCTURE analysis, we set 200,000 iterations of MCMC with sampling interval of 100, and the first half were discarded as burn-in.

We conducted chromosome painting of the MLST data using the strains sequenced in this study as donors and the known 281 STs as recipients. We used the software ChromoPainter to paint the seven genes separately with the parameter $k = 500$, and then manually combined the results together.

**Acknowledgements**

We thank Hin-chung Wong and Biao Kan for providing the strains of VP, and we thank Ichizo Kobayashi, Daniel Lawson and Julian Parkhill for valuable comments. This work was supported by the National Key Program for Infectious Diseases of China (grant number 2012ZX10004215, 2013ZX10004216 and 2013ZX10004221-002), and International Science & Technology Cooperation Program of China (grant number 2011DFA33220).

**References**

Achtman M, Wagner M 2008. Microbial diversity and the genetic nature of microbial species. Nat Rev Microbiol 6: 431-440. doi: 10.1038/nrmicro1872

Barrett JC, Fry B, Maller J, Daly MJ 2005. Haploview: analysis and visualization of LD and haplotype maps. Bioinformatics 21: 263-265. doi: 10.1093/bioinformatics/bth457

Boucher Y, Cordero OX, Takemura A, Hunt DE, Schliep K, Bapteste E, Lopez P, Tarr CL, Polz MF 2011. Local mobile gene pools rapidly cross species boundaries to create endemicity within global Vibrio cholerae populations. MBio 2. doi: 10.1128/mBio.00335-10

Castillo-Ramírez S, Harris SR, Holden MT, He M, Parkhill J, Bentley SD, Feil EJ 2011. The impact of recombination on dN/dS within recently emerged bacterial clones. PLoS pathogens 7: e1002129.

Charlesworth B 2009. Fundamental concepts in genetics: effective population size and patterns of molecular evolution and variation. Nat Rev Genet 10: 195-205. doi: 10.1038/nrg2526

Charlesworth D 2006. Balancing selection and its effects on sequences in nearby genome regions. PLoS Genet 2: e64. doi: 10.1371/journal.pgen.0020064

Cohan FM, Koeppel AF 2008. The origins of ecological diversity in prokaryotes. Curr Biol 18: R1024-1034. doi: 10.1016/j.cub.2008.09.014

Croucher NJ, Harris SR, Fraser C, Quail MA, Burton J, van der Linden M, McGee L, von Gottberg A, Song JH, Ko KS, Pichon B, Baker S, Parry CM, Lambertsen LM, Shahinas D, Pillai DR, Mitchell TJ, Dougan G, Tomasz A, Klugman KP, Parkhill J, Hanage WP, Bentley SD 2011. Rapid pneumococcal evolution in response to clinical interventions. Science 331: 430-434. doi: 10.1126/science.1198545

Cui Y, Yu C, Yan Y, Li D, Li Y, Jombart T, Weinert LA, Wang Z, Guo Z, Xu L, Zhang Y, Zheng H, Qin N, Xiao X, Wu M, Wang X, Zhou D, Qi Z, Du Z, Wu H, Yang X, Cao H, Wang H, Wang J, Yao S, Rakin A, Falush D, Balloux F, Achtman M, Song Y, Yang R 2013. Historical variations in mutation rate in an epidemic pathogen, Yersinia pestis. Proc Natl Acad Sci U S A 110: 577-582. doi: 10.1073/pnas.1205750110

Delcher AL, Salzberg SL, Phillippy AM 2003. Using MUMmer to identify similar regions in large sequence sets. Curr Protoc Bioinformatics Chapter 10: Unit 10 13. doi: 10.1002/0471250953.bi1003s00

Didelot X, Bowden R, Street T, Golubchik T, Spencer C, McVean G, Sangal V, Anjum MF, Achtman M, Falush D, Donnelly P 2011. Recombination and population structure in Salmonella enterica. PLoS Genet 7: e1002191. doi: 10.1371/journal.pgen.1002191

Didelot X, Falush D 2007. Inference of bacterial microevolution using multilocus sequence data. Genetics 175: 1251-1266. doi: 10.1534/genetics.106.063305

Didelot X, Meric G, Falush D, Darling AE 2012. Impact of homologous and non-homologous recombination in the genomic evolution of Escherichia coli. BMC Genomics 13: 256. doi: 10.1186/1471-2164-13-256

Didelot X, Nell S, Yang I, Woltemate S, van der Merwe S, Suerbaum S 2013. Genomic evolution and transmission of Helicobacter pylori in two South African families. Proc Natl Acad Sci U S A 110: 13880-13885.

Donati C, Hiller NL, Tettelin H, Muzzi A, Croucher NJ, Angiuoli SV, Oggioni M, Dunning Hotopp JC, Hu FZ, Riley DR, Covacci A, Mitchell TJ, Bentley SD, Kilian M, Ehrlich GD, Rappuoli R, Moxon ER,

Masignani V 2010. Structure and dynamics of the pan-genome of Streptococcus pneumoniae and closely related species. Genome Biol 11: R107. doi: 10.1186/gb-2010-11-10-r107

Eldon B, Wakeley J 2006. Coalescent Processes When the Distribution of Offspring Number Among Individuals Is Highly Skewed. Genetics 172: 2621-2633. doi: 10.1534/genetics.105.052175

Falush D, Stephens M, Pritchard JK 2007. Inference of population structure using multilocus genotype data: dominant markers and null alleles. Molecular Ecology Notes 7: 574-578.

Falush D, Wirth T, Linz B, Pritchard JK, Stephens M, Kidd M, Blaser MJ, Graham DY, Vacher S, Perez-Perez GI, Yamaoka Y, Megraud F, Otto K, Reichard U, Katzowitsch E, Wang X, Achtman M, Suerbaum S 2003. Traces of human migrations in Helicobacter pylori populations. Science 299: 1582-1585. doi: 10.1126/science.1080857

Ferreira RB, Chodur DM, Antunes LC, Trimble MJ, McCarter LL 2012. Output targets and transcriptional regulation by a cyclic dimeric GMP-responsive circuit in the Vibrio parahaemolyticus Scr network. J Bacteriol 194: 914-924. doi: 10.1128/JB.05807-11

Fraser C, Alm EJ, Polz MF, Spratt BG, Hanage WP 2009. The bacterial species challenge: making sense of genetic and ecological diversity. Science 323: 741-746.

Fraser C, Hanage WP, Spratt BG 2005. Neutral microepidemic evolution of bacterial pathogens. Proc Natl Acad Sci U S A 102: 1968-1973. doi: 10.1073/pnas.0406993102

Fujino T, Miwatani T, Yasuda J, Kondo M, Takeda Y, Akita Y, Kotera K, Okada M, Nishimune H, Shimizu Y, Tamura T, Tamura Y 1965. Taxonomic studies on the bacterial strains isolated from cases of "shirasu" food-poisoning (Pasteurella parahaemolytica) and related microorganisms. Biken J 8: 63-71.

Gomelsky M 2012. Cyclic Dimeric GMP-Mediated Decisions in Surface-Grown Vibrio parahaemolyticus: a Different Kind of Motile-to-Sessile Transition. J Bacteriol 194: 911-913. doi: 10.1128/JB.06695-11

Gressmann H, Linz B, Ghai R, Pleissner KP, Schlapbach R, Yamaoka Y, Kraft C, Suerbaum S, Meyer TF, Achtman M 2005. Gain and loss of multiple genes during the evolution of Helicobacter pylori. PLoS Genet 1: e43. doi: 10.1371/journal.pgen.0010043

Grissa I, Vergnaud G, Pourcel C 2007. The CRISPRdb database and tools to display CRISPRs and to generate dictionaries of spacers and repeats. BMC Bioinformatics 8: 172. doi: 10.1186/1471-2105-8-172

Guindon S, Gascuel O 2003. A simple, fast, and accurate algorithm to estimate large phylogenies by maximum likelihood. Syst Biol 52: 696-704. doi: 54QHX07WB5K5XCX4 [pii]

Han H, Wong HC, Kan B, Guo Z, Zeng X, Yin S, Liu X, Yang R, Zhou D 2008. Genome plasticity of Vibrio parahaemolyticus: microevolution of the 'pandemic group'. BMC Genomics 9: 570. doi: 10.1186/1471-2164-9-570

Hyatt D, Chen GL, Locascio PF, Land ML, Larimer FW, Hauser LJ 2010. Prodigal: prokaryotic gene recognition and translation initiation site identification. BMC Bioinformatics 11: 119. doi: 10.1186/1471-2105-11-119

Jolley K, Wilson D, Kriz P, McVean G, Maiden M 2005. The influence of mutation, recombination, population history, and selection on patterns of genetic diversity in Neisseria meningitidis. Mol Biol Evol 22: 562-569.

Kennemann L, Didelot X, Aebischer T, Kuhn S, Drescher B, Droege M, Reinhardt R, Correa P, Meyer TF, Josenhans C 2011. Helicobacter pylori genome evolution during human infection. Proc Natl Acad Sci U S A 108: 5033-5038.

Kulick S, Moccia C, Didelot X, Falush D, Kraft C, Suerbaum S 2008. Mosaic DNA imports with interspersions of recipient sequence after natural transformation of Helicobacter pylori. PLoS One 3: e3797. doi: 10.1371/journal.pone.0003797

Lawson DJ, Falush D 2012. Population identification using genetic data. Annu Rev Genomics Hum Genet 13: 337-361. doi: 10.1146/annurev-genom-082410-101510

Lawson DJ, Hellenthal G, Myers S, Falush D 2012. Inference of population structure using dense haplotype data. PLoS Genet 8: e1002453. doi: 10.1371/journal.pgen.1002453

Letunic I, Doerks T, Bork P 2014. SMART: recent updates, new developments and status in 2015. Nucleic Acids Res. doi: 10.1093/nar/gku949

Li H, Durbin R 2011. Inference of human population history from individual whole-genome sequences. Nature 475: 493-496. doi: 10.1038/nature10231

Li R, Zhu H, Ruan J, Qian W, Fang X, Shi Z, Li Y, Li S, Shan G, Kristiansen K, Yang H, Wang J 2010. De novo assembly of human genomes with massively parallel short read sequencing. Genome Res 20: 265-272. doi: 10.1101/gr.097261.109

Marchler-Bauer A, Zheng C, Chitsaz F, Derbyshire MK, Geer LY, Geer RC, Gonzales NR, Gwadz M, Hurwitz DI, Lanczycki CJ, Lu F, Lu S, Marchler GH, Song JS, Thanki N, Yamashita RA, Zhang D, Bryant SH 2013. CDD: conserved domains and protein three-dimensional structure. Nucleic Acids Res 41: D348-352. doi: 10.1093/nar/gks1243

Martiny JB, Bohannan BJ, Brown JH, Colwell RK, Fuhrman JA, Green JL, Horner-Devine MC, Kane M, Krumins JA, Kuske CR, Morin PJ, Naeem S, Ovreas L, Reysenbach AL, Smith VH, Staley JT 2006. Microbial biogeography: putting microorganisms on the map. Nat Rev Microbiol 4: 102-112. doi: 10.1038/nrmicro1341

Matic I, Radman M, Rayssiguier C 1994. Structure of recombinants from conjugational crosses between Escherichia coli donor and mismatch-repair deficient Salmonella typhimurium recipients. Genetics 136: 17-26.

Maynard Smith J, Smith NH, O'Rourke M, Spratt BG 1993. How clonal are bacteria? Proc Natl Acad Sci U S A 90: 4384-4388.

McCarthy ND, Colles FM, Dingle KE, Bagnall MC, Manning G, Maiden MC, Falush D 2007. Host-associated genetic import in Campylobacter jejuni. Emerg Infect Dis 13: 267-272. doi: 10.3201/eid1302.060620

Moodley Y, Linz B, Yamaoka Y, Windsor HM, Breurec S, Wu J-Y, Maady A, Bernhöft S, Thiberge J-M, Phuanukoonnon S 2009. The peopling of the Pacific from a bacterial perspective. Science 323: 527-530.

Nair GB, Ramamurthy T, Bhattacharya SK, Dutta B, Takeda Y, Sack DA 2007. Global dissemination of Vibrio parahaemolyticus serotype O3:K6 and its serovariants. Clin Microbiol Rev 20: 39-48. doi: 10.1128/CMR.00025-06

Okuda J, Ishibashi M, Hayakawa E, Nishino T, Takeda Y, Mukhopadhyay AK, Garg S, Bhattacharya SK, Nair GB, Nishibuchi M 1997. Emergence of a unique O3:K6 clone of Vibrio parahaemolyticus in Calcutta, India, and isolation of strains from the same clonal group from Southeast Asian travelers arriving in Japan. J Clin Microbiol 35: 3150-3155.

Park L 2012. Linkage disequilibrium decay and past population history in the human genome. PLoS One 7: e46603. doi: 10.1371/journal.pone.0046603

Pritchard JK, Stephens M, Donnelly P 2000. Inference of population structure using multilocus

genotype data. Genetics 155: 945-959.

Rocha EP, Smith JM, Hurst LD, Holden MT, Cooper JE, Smith NH, Feil EJ 2006. Comparisons of dN/dS are time dependent for closely related bacterial genomes. J Theor Biol 239: 226-235. doi: 10.1016/j.jtbi.2005.08.037

Rosenberg NA 2004. DISTRUCT: a program for the graphical display of population structure. Molecular Ecology Notes 4: 137-138.

Salomon D, Gonzalez H, Updegraff BL, Orth K 2013. Vibrio parahaemolyticus type VI secretion system 1 is activated in marine conditions to target bacteria, and is differentially regulated from system 2. PLoS One 8: e61086. doi: 10.1371/journal.pone.0061086

Shapiro BJ, Friedman J, Cordero OX, Preheim SP, Timberlake SC, Szabo G, Polz MF, Alm EJ 2012. Population genomics of early events in the ecological differentiation of bacteria Science 336: 48-51. doi: 10.1126/science.1218198

Shen P, Huang HV 1986. Homologous recombination in Escherichia coli: dependence on substrate length and homology. Genetics 112: 441-457.

Sheppard SK, McCarthy ND, Falush D, Maiden MC 2008. Convergence of Campylobacter species: implications for bacterial evolution. Science 320: 237-239. doi: 10.1126/science.1155532

Su YC, Liu C 2007. Vibrio parahaemolyticus: a concern of seafood safety. Food Microbiol 24: 549-558. doi: 10.1016/j.fm.2007.01.005

Tenaillon O, Skurnik D, Picard B, Denamur E 2010. The population genetics of commensal Escherichia coli. Nat Rev Microbiol 8: 207-217. doi: 10.1038/nrmicro2298

Tischler AD, Camilli A 2004. Cyclic diguanylate (c-di-GMP) regulates Vibrio cholerae biofilm formation. Mol Microbiol 53: 857-869. doi: 10.1111/j.1365-2958.2004.04155.x

Voight BF, Kudaravalli S, Wen X, Pritchard JK 2006. A map of recent positive selection in the human genome. PLoS Biol 4: e72. doi: 10.1371/journal.pbio.0040072

Wang J 2005. Estimation of effective population sizes from data on genetic markers. Philos Trans R Soc Lond B Biol Sci 360: 1395-1409. doi: 10.1098/rstb.2005.1682

Whitlock MC, McCauley DE 1999. Indirect measures of gene flow and migration: $F_{ST} \neq 1/(4Nm+1)$. Heredity 82: 117-125.

Yahara K, Furuta Y, Oshima K, Yoshida M, Azuma T, Hattori M, Uchiyama I, Kobayashi I 2013. Chromosome painting in silico in a bacterial species reveals fine population structure. Mol Biol Evol 30: 1454-1464. doi: 10.1093/molbev/mst055

Yan Y, Cui Y, Han H, Xiao X, Wong HC, Tan Y, Guo Z, Liu X, Yang R, Zhou D 2011. Extended MLST-based population genetics and phylogeny of Vibrio parahaemolyticus with high levels of recombination. Int J Food Microbiol 145: 106-112. doi: 10.1016/j.ijfoodmicro.2010.11.038

Yang Z 2007. PAML 4: phylogenetic analysis by maximum likelihood. Mol Biol Evol 24: 1586-1591. doi: 10.1093/molbev/msm088

Yeung PS, Boor KJ 2004. Epidemiology, pathogenesis, and prevention of foodborne Vibrio parahaemolyticus infections. Foodborne Pathog Dis 1: 74-88. doi: 10.1089/153531404323143594

Yu Y, Yang H, Li J, Zhang P, Wu B, Zhu B, Zhang Y, Fang W 2012. Putative type VI secretion systems of Vibrio parahaemolyticus contribute to adhesion to cultured cell monolayers. Arch Microbiol 194: 827-835. doi: 10.1007/s00203-012-0816-z

Zhou Z, McCann A, Weill FX, Blin C, Nair S, Wain J, Dougan G, Achtman M 2014. Transient Darwinian selection in Salmonella enterica serovar Paratyphi A during 450 years of global spread of enteric fever. Proc Natl Acad Sci U S A. doi: 10.1073/pnas.1411012111

**Figure legends**

**Figure 1 Neighbor-joining tree of 157 VP strains constructed using 327,904 SNPs.** The different colors from outer to inner indicate the isolated location, time, epistasis groups and source for each strain, respectively. The blank indicated the data was not available, or minor groups defined by the epistasis locus (LS001). Branch color indicate the sub-populations defined by fineSTRUCTURE.

**Figure 2 Phylogeny of pandemic strains and the SNP distribution across chromosome in different sets of genomes.** (A) Maximum likelihood tree of pandemic strains based on non-recombinant SNPs. The scale bar represents substitutions per SNP. The tree was rooted using strain S093. There were 97.3% of recombinant sites concentrated in three branches (indicated by grey triangles) leading to different serovariants of pandemic strains. The grey circle indicates the sublineage that only contained strains isolated from Taiwan, China. (B-E) The distributions of SNPs were plotted for strains including (B) pandemic strains CG1; (C) CG1 and S093; (D) CG1, S093 and CG2; and (E) CG1, S093, CG2 and S002 (randomly selected from the remaining strains). The number of SNPs were counted within 10 kb windows across the chromosome. The grey shading indicates the recombination hot region surrounding the O- and K- antigen encoding gene cluster.

**Figure 3 Population structure according to fineSTRUCTURE.** (A) Coancestry matrix constructed from 71 strains. (B) Coancestry matrix based on all strains of Asia-Pop 1 and one strain from each of the remaining sub-populations. The colors indicate the coancestry values

of population averages representing chunk numbers from donors (column) to recipients (row), and the trees show the sub-populations and their clustering relative to each other.

**Figure 4 Population structure of VP based on MLST.** (A) Population structure of 281 sequence types (STs) of VP inferred by STRUCTURE with K = 10. Each bar represents an ST, and the colors indicate the proportion of ancestral sequences inherited from each of the nine hypothetical ancestral populations. (B) Coancestry heatmap inferred by fineSTRUCTURE based on the MLST genes. The columns represent STs ordered as in (A) and the rows represent sub-populations inferred by the fineSTRUCTURE analysis based on genome sequences. The color in each cell represents the population average number of chunks from donor (strains that were sequenced in this research) to recipient (STs from public MLST database). (C) Geographical source of VP strains in each STs. The Atlantic, Pacific and Mexican Gulf categories indicate strains isolated from states of the east and west coast of USA, and from states surrounding the Mexican Gulf. The USA category indicate inland US states or no state information.

**Figure 5 Effective population size of bacteria.** (A) Clonal genealogy and genealogies for individual sites. The genealogies coincide until the most recent homologous recombination rate at each site. (B) Clonal genealogy under an ecotype model. Ecotypes persist stably together in the population once they have evolved, so that coalescences between strains in the sample that have different ecotypes happen at a low rate.

**Figure 6 Estimation of the effective population size of the Asian population.** (A)

Correlation between the number of new mutations within clonal lineages and the total size of recombined regions within the same lineage. Each dot indicated one clonal lineage. The linear regression was constrained to go through the origin and the grey shading indicated the 99 CI% of the slope. (B) Distances in a phylogenetic tree. The red dots represent the distances of ancestors in a UPGMA phylogeny of the Asian genomes, and the blue curve indicates the corresponding expectation for a sample of size *n*=60 and parameter $N_e r = 268$. From left to right, the dot curves indicated the expectation values when $N_e r$ were 10,50,150,350,450, separately.

**Figure 7 Q-Q plot showing the distribution of *P* values for associations between pairs of SNPs (A) and between SNPs and flexible genome blocks (B).** The x-axis indicates the *P* value and the y-axis indicates the frequency of corresponding *P* value. Both axes are on a logarithmic scale. The red circles are based on all tested pairs and the blue triangles included only pairs of SNPs (panel A) or pairs between SNP and flexible genomes (panel B) with distance larger than 3 kb. The black line is based on a linear regression of the data points with log (*P* value) larger than -6. The red line is based on a linear regression of the red circles that range from -14 to -8 of log (*P* value). The slope of the blue line in panel A is the same as the red one, and its intercept was generated by least squares fitting to the blue triangles with log (*P* value) smaller than -8.

**Figure 8 Position of the epistasis loci in the reference genome.** The semi-circles indicated two chromosomes of the VP. The red bars indicated the SNPs and blue ones are flexible genome blocks. As the flexible genome block that carried cellulose biosynthesis proteins was plot independently as being absent in the reference genome. The black lines that linked

epistasis loci indicated the strong association present amongst them with $P < 10^{-6}$.

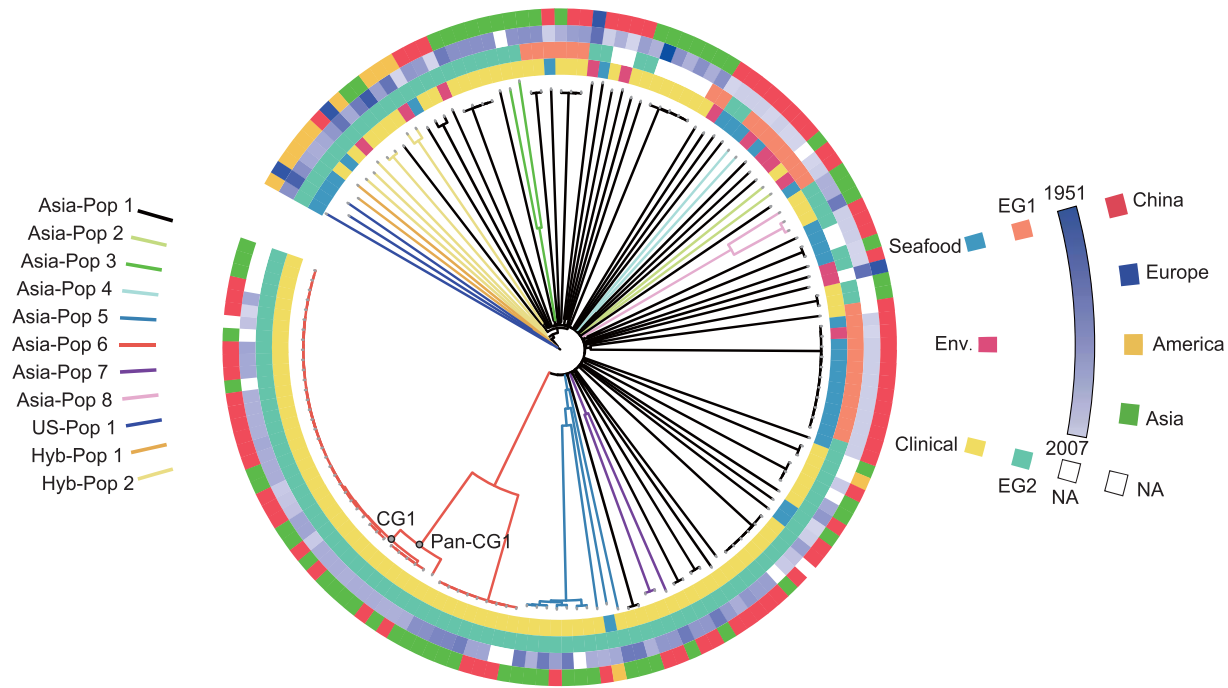

Figure 1

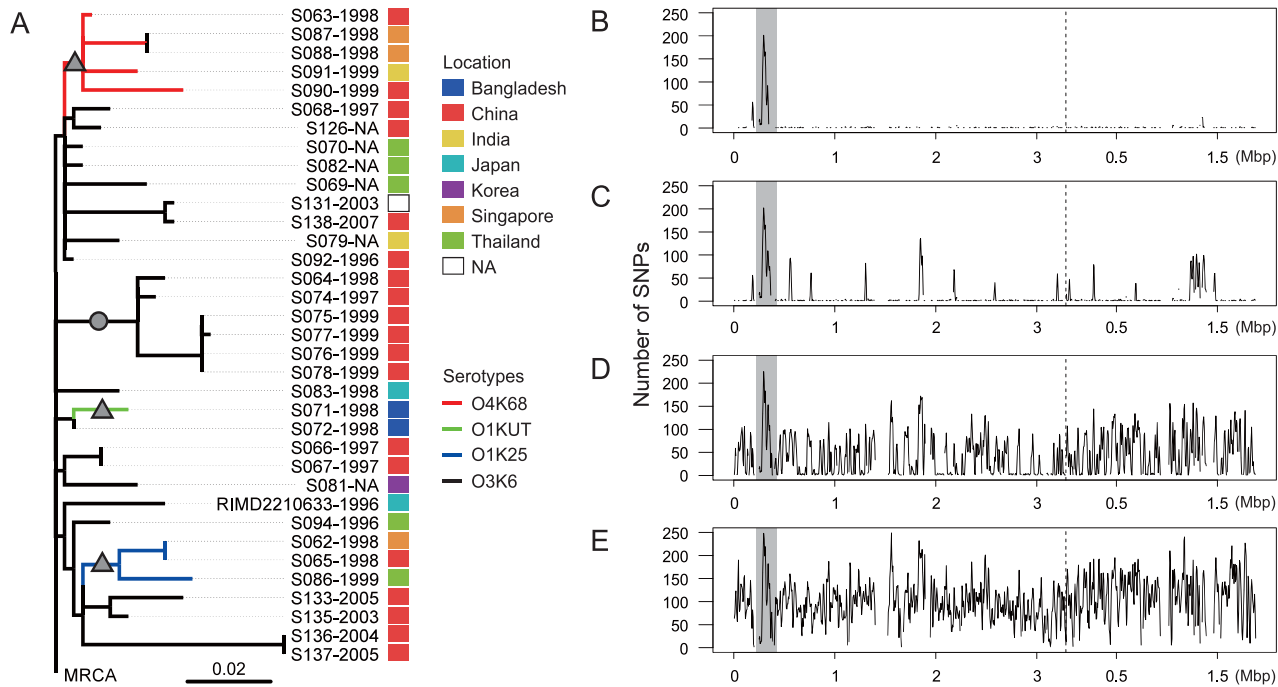

**Figure 2**

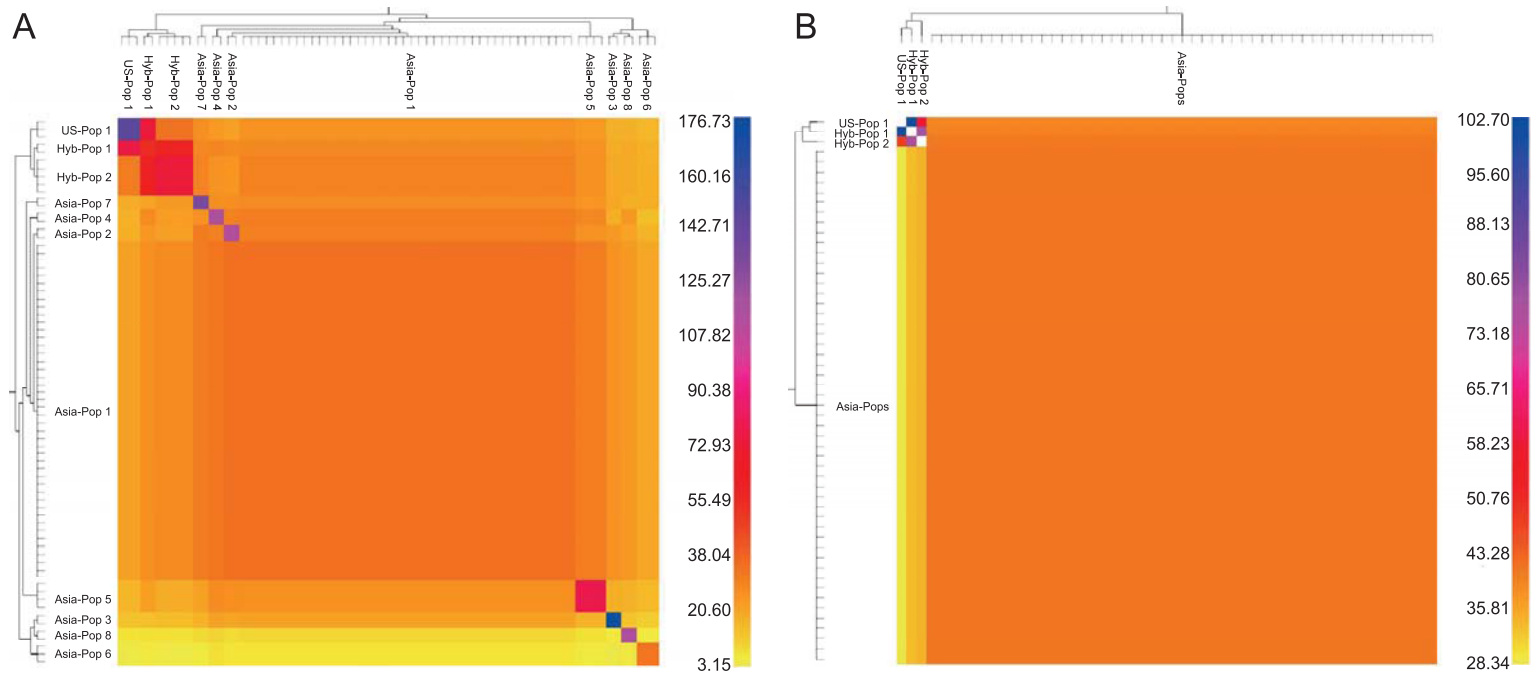

**Figure 3**

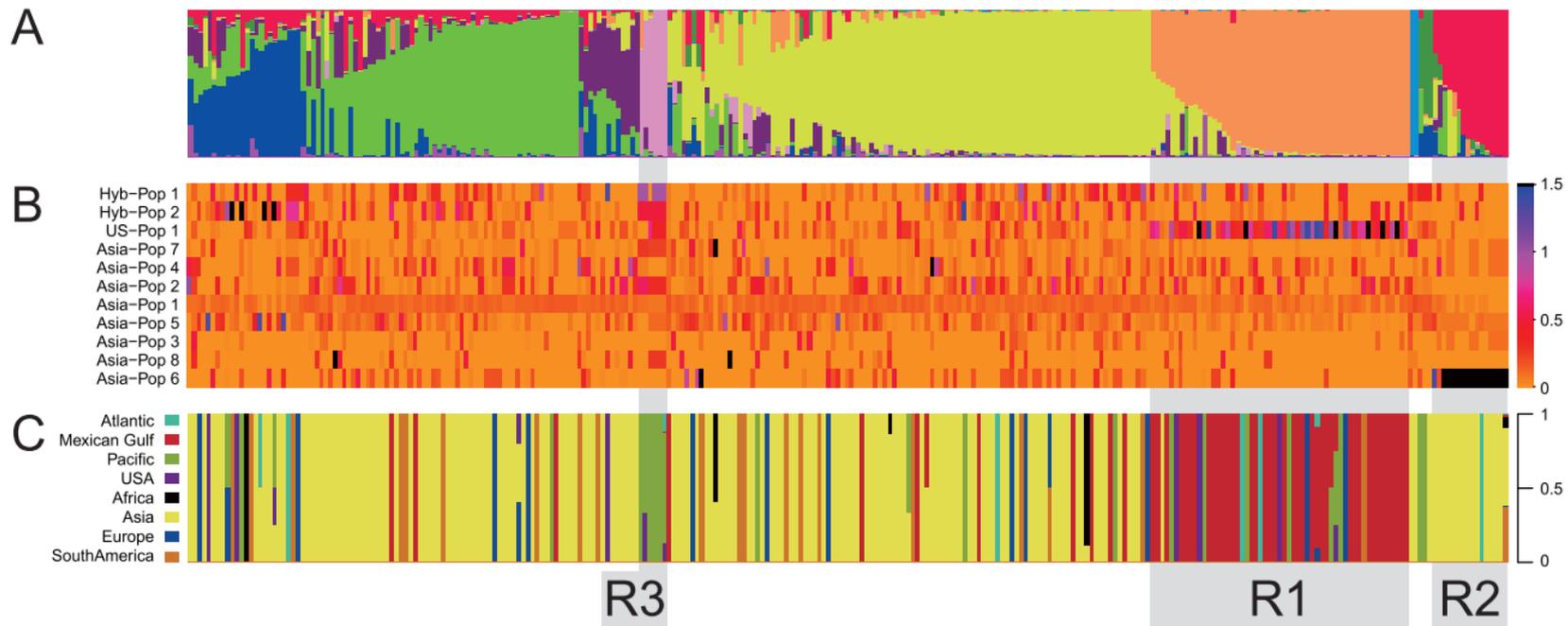

**Figure 4**

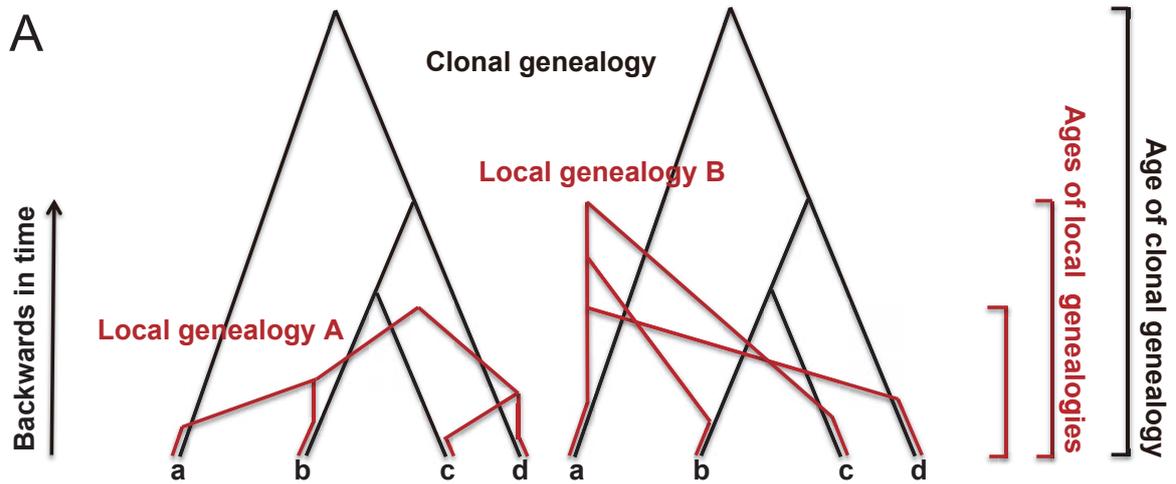
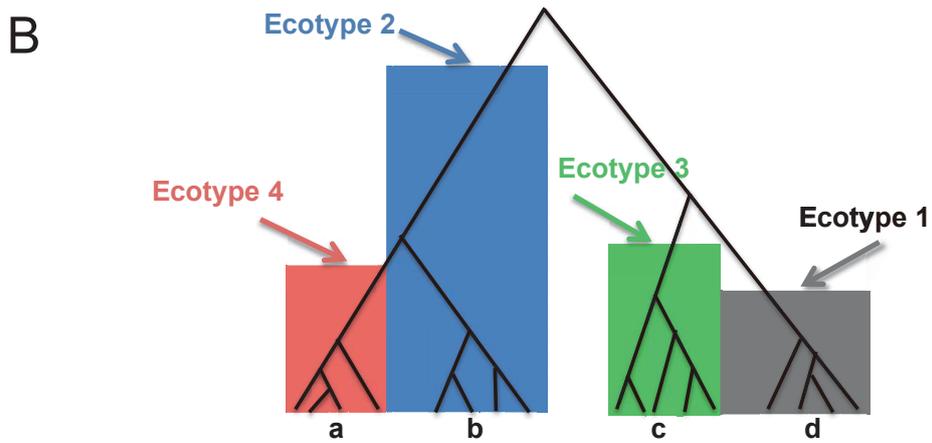

**Figure 5**

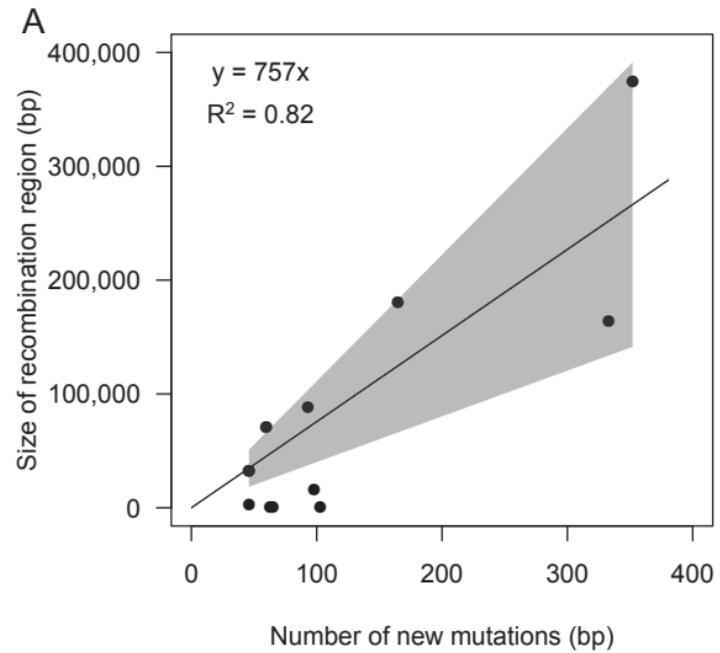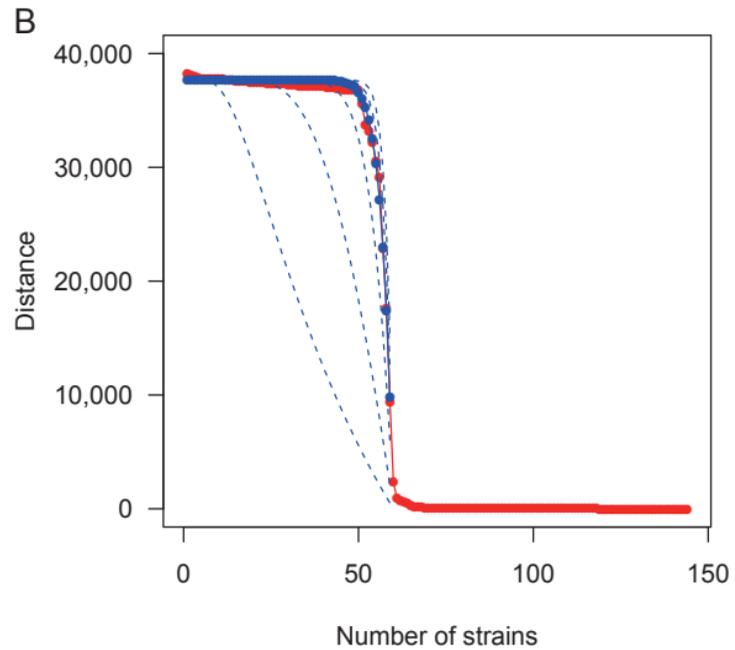

**Figure 6**

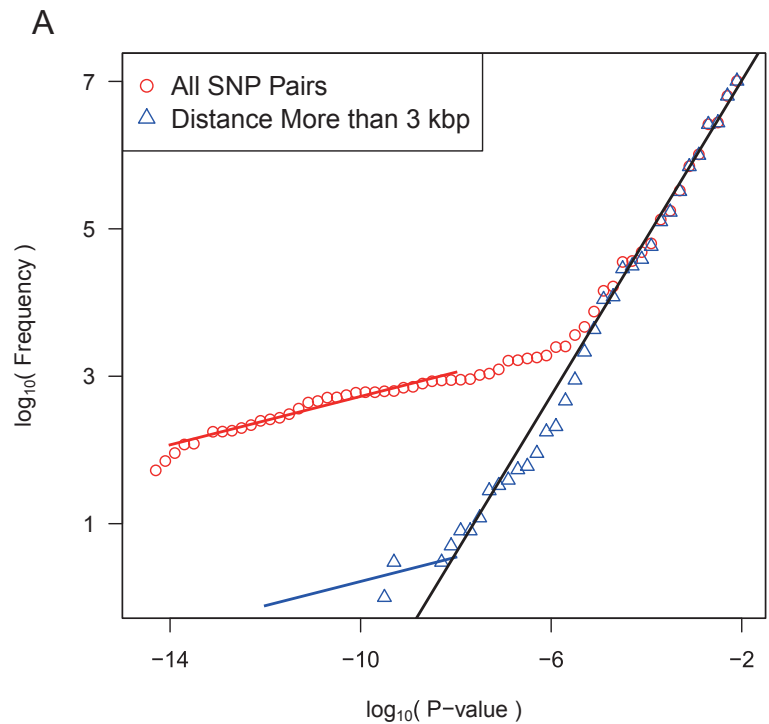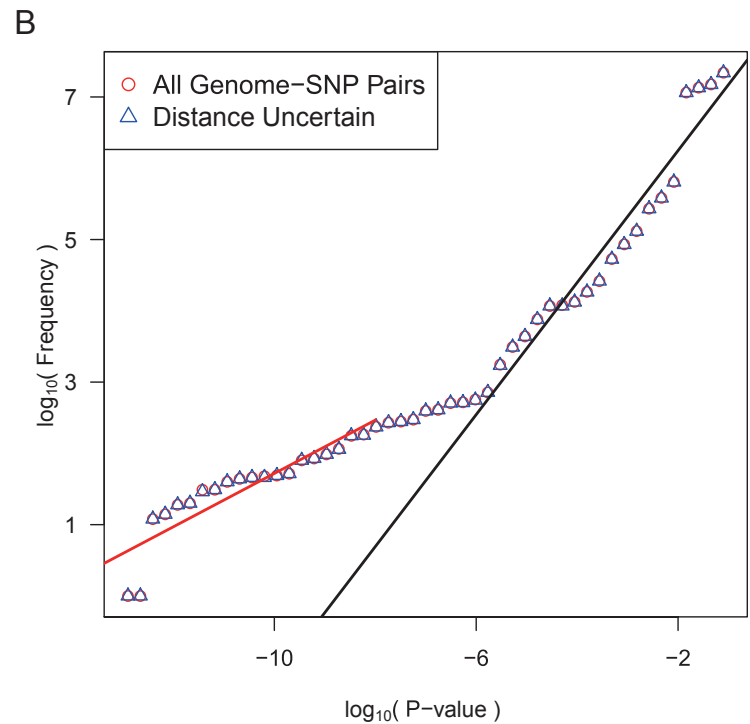

**Figure 7**

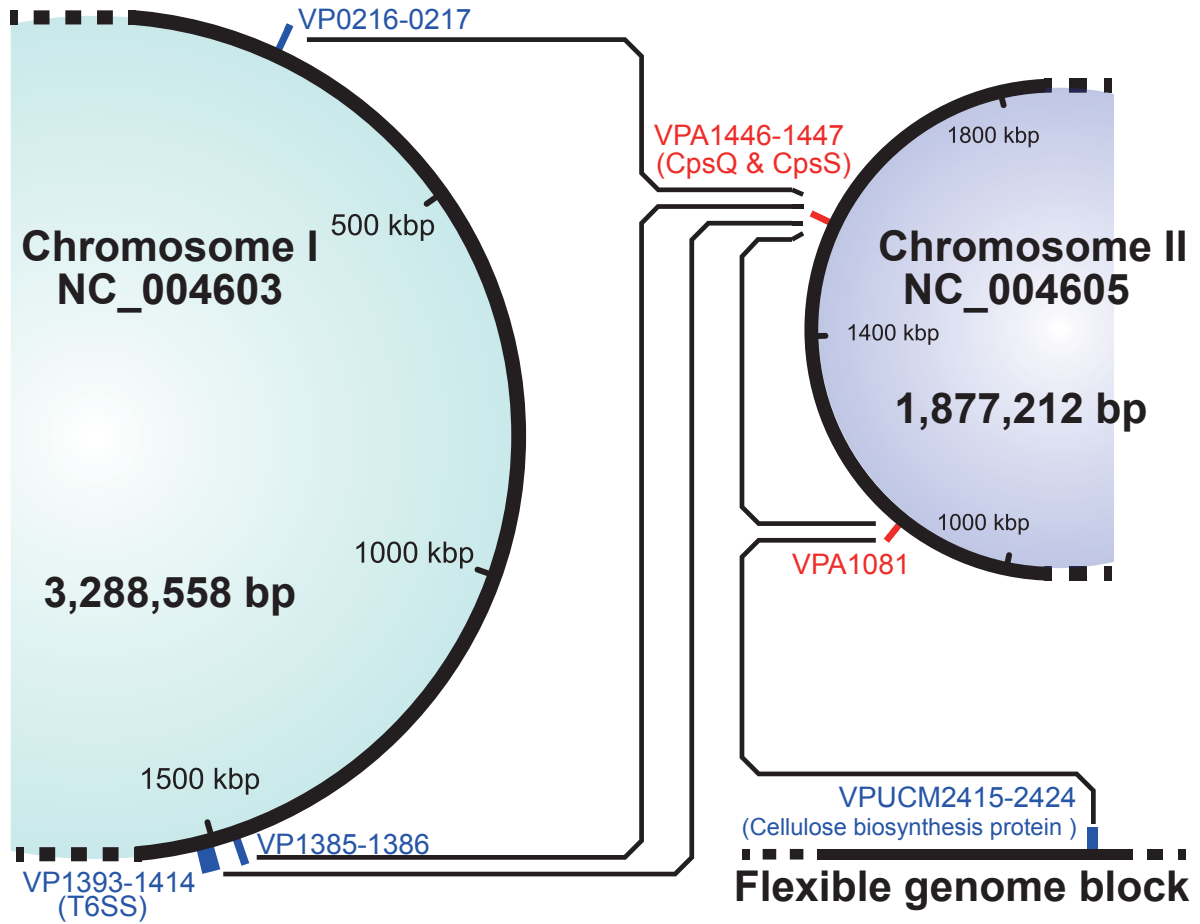

**Figure 8**